\documentclass[prb,amsmath,superscriptaddress,citeautoscript,twocolumn,showpacs,floatfix]{revtex4}

\usepackage{bm}
\usepackage{amssymb}
\usepackage{graphicx}
\usepackage{graphicx}

\newcommand{\refcite}{\onlinecite}

\begin{document}

\title[The localization transition at finite temperatures: electric and thermal transport]{The localization transition at finite temperatures: electric and thermal transport \\
 {\normalsize \textbf{Article for:  "50 Years of Anderson Localization"}}}

\author{Yoseph Imry}
\email{Yoseph.Imry@weizmann.ac.il}

\author{Ariel Amir}

\affiliation{Department of Condensed Matter Physics,  Weizmann
Institute of Science, Rehovot 76100, Israel}

\date{\today}

\begin{abstract}

The Anderson localization transition is considered at finite
temperatures. This includes the electrical conductivity as well as
the electronic thermal conductivity and the thermoelectric
coefficients. An interesting critical behavior  of the latter is
found. A method for characterizing the conductivity critical exponent, an
important signature of the transition, using the conductivity and
thermopower measurements, is outlined.

\end{abstract}



\maketitle

\def\ul{\underline}
\def\la{\langle}
\def\ra{\rangle}
\def\be{\begin{equation}}
\def\ee{\end{equation}}
\section{Introduction}\label{sec1.1}

Anderson localization \cite{Anderson,Mott-B} is a remarkable, and
very early, example of a quantum phase transition (QPT), where the
nature of a system at $T = 0$ changes abruptly and nonanalytically
at a point, as a function of a control parameter. Here it implies
that the relevant quantum states of the system acquire an
exponential decay at large distances (similar to, but much more
complex than, the formation of a bound state). In the original paper
this phenomenon was discovered via a change in the convergence
properties of the ``locator" expansion. In the disordered
tight-binding model, with short-range hopping, this is the expansion
about the bound atomic, or Wannier-type orbitals. As long as the
expansion converges, the relevant eigenstates are localized; its
divergence signifies the transition to ``extended", delocalized
states. This was reviewed, hopefully pedagogically, in Ref.
[\onlinecite{Avi}]. Later, Mott introduced the very useful picture
of the ``mobility edge" within the band of allowed energies,
separating localized from extended states \cite{Mott-B}. In  the
lower part of the band, the states below the ``lower mobility edge"
are localized, while those above it are extended. When the disorder
and/or the position of the Fermi level, $E_{F}$, are changed, a
point where they cross each other is where the states at $E_{F}$
change nature from extended to localized, and this is a simple and
instructive model for a  metal-insulator transition at vanishing
temperature $T$.

The analysis of the above localization transition centers on the
behavior of $\sigma_0(E)$, the conductivity at energy E, which would
be the $T = 0$ conductivity of the sample with $E_{F} = E$.
$\sigma_0(E)$ vanishes for $E$ below the lower mobility edge $E_{m}$
and goes to zero when $E$ approaches $E_{m}$ from above \cite{Gang4}
\be \sigma_0(E)= A (E - E_{m})^{x}. \label{crit1} \ee The
characteristic exponent  for this, $x$, is an important parameter of
the theory. It is expected to be universal for a large class of
noninteracting models, but its value is not really known, in spite
of the several analytical, numerical and experimental methods used to attempt its
evaluation.

The electron-electron interaction is certainly relevant near this
transition, which may bring it to a different universality class.
This is a difficult problem. The benchmark treatment is the one by
Finkel'stein \cite{Sasha}. Since it is likely that the situation in
at least most experimental systems is within this class, it would
appear impossible to determine the value of $x$ for the pure
Anderson transition (without interactions)\cite{small}.

In this paper we still analyze the thermal and thermoelectric transport for a general
model with $\sigma_0(E)$ behaving as in Eq. (\ref{crit1}). This is
certainly valid for noninteracting electrons and should be valid
also including the interactions, as long as some kind of Landau
Fermi-liquid quasiparticles exist. In that case $\sigma_0(E)$ for
quasiparticles is definable and thermal averaging with Fermi
statistics holds. Even then, however, various parameter
renormalizations and interaction  corrections \cite{AA} should come
in. The effect of the interactions on the thermopower for a small
system in the Coulomb blockade picture was considered in Ref.
[\onlinecite{Carlo}], and
 correlations were included in Ref. [\onlinecite{Fl}]. Even in the latter case, the simple Cutler-Mott \cite{CM} formula
 (Eq. \ref{S}, derived in section \ref{therm}) was found to work surprisingly well.

It should be mentioned that the sharp and asymmetric energy-dependence of
$\sigma_0(E)$ near the mobility edge \cite{CM,SI} should and
does \cite{CL,CM,Zvi} lead  to rather large values
of the thermopower. Exceptions will be mentioned and briefly discussed later.
Large thermopowers are important for energy conversion and refrigeration
applications \cite{HH} and this clearly deserves further studies.

A serious limitation on the considerations presented here is that
the temperature should be low enough so that all the inelastic
scattering (electron-phonon, electron-electron, etc.) is negligible.
For simplicity we consider here {\em only} longitudinal transport
(currents parallel to the driving fields). Thus, no Hall  or
Nernst-Ettinghausen effect! We also do not consider the
thermoelectric transport in the hopping regime. It might be relevant
even in the metallic regime (chemical potential  $\mu > E_{m}$),
once $k_{B} T \gtrsim \mu - E_{m}$. This may well place limitations
on the high temperature analysis we make in the following.

In section \ref{scal} we review the
basic concepts behind the scaling theory \cite{Gang4} for the
transition, and reiterate the critical behavior of the conductivity
as in Eq. (\ref{crit1}), obtaining also its temperature dependence.
In section \ref{therm} we derive all results for the thermal and
thermoelectric transport and analyze the scaling critical behavior
of the latter as function of temperature and distance from the
transition. A brief comparison with experiment is done in section
\ref{exp} and concluding remarks are given in section \ref{Disc}. In the appendix
we present a proof that the heat carried
by a quasiparticle is equal to its energy measured from the chemical potential, $\mu$, hoping that
this also clarifies the physics of  this result.  This
derivation is valid for Bose quasiparticles (phonons, photons, magnons, excitons, plasmons, etc.) as well.

\section{The zero and finite temperature macroscopic conductivity around the Anderson localization transition} \label{scal}

\subsection{The Thouless picture within the tunnel-junction model}
We start this section by briefly reviewing the tunnel-junction picture of
conduction \cite{Bardeen,Harrison} at $T = 0$, which is a useful way
to understand the important  Thouless \cite{Thouless} picture for
such transport. Consider first two pieces (later referred to as
``blocks") of a conducting material with a linear size $L$,
connected through a layer of insulator (usually an oxide) which is
thin enough to allow for electron tunneling. The interfaces are
assumed rough, so there is no conservation of the transverse
momentum: each state on the left interacts with each state on the
right with a matrix element $t$ with a roughly uniform absolute
value. The lifetime $\tau_{_L}$  for an electron on one such block
for a transition to the other one is given by the  Fermi golden rule
(at least when tunneling is a weak perturbation): \be
\tau_{_L}^{-1}=\frac{2\pi}{\hbar}\overline{|t|^2}N_r(E_{_F}),
\label{GR} \ee where $\overline{|t|^2}$ is the average of the
tunneling matrix element squared and $N_r(E_{_F})$ is the density of
states on the final (right-hand) side. Taking the density of states
(DOS) in the initial side to be $N_\ell(E_{_F})$, we find that when
a voltage V is applied, $eVN_\ell(E_{_F})$ states are available,
each decaying to the right with a time constant $\tau_{_L}$, so that
the current is \be I=e^2 N_\ell(E_{_F})\tau^{-1}_{_L}V,
\label{current} \ee and the conductance is \be
G=e^2N_\ell(E_{_F})/\tau_{_L}=\frac{2\pi e^2}{\hbar}\overline{t^2}
N_\ell(E_{_F})N_r(E_{_F})\, ,\label{G_current} \ee which is an
extremely useful result. This equality is well-known in the tunnel
junction theory \cite{Bardeen,Harrison}. Clearly, G is symmetric
upon exchanging $l$ and $r$, as it should \cite{lrsymm}. Note that
Eqs. (\ref{GR}) and (\ref{G_current}) are valid in any number of
dimensions. An important remark is that Eq. (\ref{GR}) necessitates
a continuum of final states, while the final (RHS) block is finite
and has a discrete spectrum. One must make the assumption that the
interaction of that system with the outside world leads to a level
broadening larger than, or on the same order of, the level spacing
\cite{Cz-Kr,T-K,book}. This is the case in most mesoscopic systems.
One then naively assumes that this condition converts the spectrum
to an effectively continuous one (the situation may actually be more
subtle \cite{Ariel}). Otherwise, when levels really become discrete,
one gets into the {\it really microscopic (molecular) level}.
\par The result of Eq. (\ref{G_current}) is very general. Let us use it for the
following scaling picture  \cite{Thouless}: Divide a large sample to
(hyper) cubes or ``blocks" of side $L$. We consider the case
$L\gg\ell,a$; $\ell$ being the elastic mean free path and $a$ the
microscopic length. The typical level separation for a block at the
relevant energy  (say, the Fermi level), $d_{_L}$, is given by the
inverse of the density of states (per unit energy) for size $L$,
$N_{L}(E_{_F})$. Defining an energy associated with the transfer of
electrons between two such adjacent systems by
$V_L\equiv\pi\hbar/\tau_{_L}$ ($\tau_{_L}$ is the lifetime of an
electron on one side against transition to the other side) the
dimensionless interblock conductance $g_{_L}\equiv
G_{_L}/(e^2/\pi\hbar)$ is: \be g_{_L}=V_{_L}/d_{_L} \label{Thouless}
\ee i.e. $g_{_L}$ is the (dimensionless) ratio of the only two
relevant energies in the problem. The way Thouless argued for this
relation is by noting that the electron's diffusion on the scale $L$
is a random walk with a step $L$ and characteristic time
$\tau_{_L}$, thus
$$D_{_L}\sim L^2/\tau_{_L}$$
Note that as long as the classical diffusion picture holds, $D_L$ is
independent of $L$ and $\tau_L=L^2/D$, which is the diffusion time
across the block. It will turn out that the localization or quantum
effects, when applicable, cause $D_L$ to decrease with $L$. For
metals, the conductivity, $\sigma_{_L}$, on the scale of the block
size $L$, is given by the Einstein relation $\sigma = D_{L} e^{2}
dn/d\mu $ (where $\mu$ is the chemical potential and $dn/d\mu$ is
the density of states per unit volume), and the conductance in d
dimensions is given by:
 \be G_{_L}= \sigma_{_L} L^{d-2}. \label {G_scaling}\ee Putting these
relations together and remembering that $N_{_L}(E{_F})\sim L^d
dn/d\mu$, yields Eq. (\ref{Thouless}).

To get some physical feeling for the energy $h/\tau_{_L}$ we note
again that, at least for the weak coupling case, the Fermi golden
rule yields Eq. (\ref{GR}) or: \be
V_{_L}=2\pi^2\overline{|t|^2}/d_{_R}. \label{VL} \ee Thus, $V_{_L}$
is defined in terms of the interblock matrix elements. Clearly, when
the blocks are of the same size, Eq. (\ref{VL}) is also related to
the order of magnitude of the second order perturbation theory shift
of the levels in one block by the interaction with the other. For a
given block this is similar to a surface effect -- the shift in the
block levels due to changes in the boundary conditions on the
surface of the block. Indeed, Thouless has given appealing physical
arguments for the equivalence of $V_L$ with the sensitivity of the
block levels to boundary conditions. This should be valid for $L$
much larger than $\ell$ and all other microscopic lengths.

\par Since in this scaling picture the separations among the blocks are fictitious
for a homogeneous system, it is clear that the interblock
conductance is just the conductance of a piece whose size is of the
order of $L$, i.e., this is the same order of magnitude as the
conductance of the block itself.
\par The latter can also be calculated using the Kubo linear response expression.
It has to be emphasized that the Kubo formulation also applies
strictly only for an infinite system whose spectrum is continuous.
For a finite system, it is argued again that a very small coupling
of the electronic system to some large bath (e.g. the phonons, or to
a large piece of conducting material) is needed to broaden the
discrete levels into an effective continuum. Edwards and  Thouless
\cite{Ed-Th}, using the Kubo-Greenwood formulation, made the
previously discussed relationship of $V_L$ with the sensitivity to
boundary conditions very precise.
\par The above picture is at the basis of the finite-size scaling \cite{Gang4,K-M} theory of localization. It can also can be
used  for numerical calculations of $g(L)$, which is a most relevant
physical parameter of the problem, for non-interacting electrons, as
we shall see. Alternatively, Eq. (\ref{VL}) as well as
generalizations thereof can and have been used for numerical
computations. Powerful numerical methods exist to this end
\cite{Fi-Le}.

It is important to emphasize that $g_{_L}\gg 1$ means that states in
neighboring blocks are tightly coupled while $g_L\ll 1$ means that
the states are essentially single-block ones. $g_L$ is therefore a
good general dimensionless measure of the strength of the coupling
between two quantum systems. Thus, if $g_{_L} \gg 1$ for small $L$
and $g_{_L}\rightarrow 0$ for $L\rightarrow\infty$, then the range
of scales $L$ where $g_{_L}\sim 1$ gives the order of magnitude of
the localization length, $\xi$.
\par Although the
above analysis was done specifically for non-interacting electrons,
it is of greater generality. $g_L$ (with obvious factors) may play
the role of a conductance also when a more general entity (e.g. an
electron pair) is transferred between the two blocks. The real
limitations for the validity of this picture seem to be  the
validity of the Fermi-liquid picture and that no inelastic effects
(e.g with phonons or electron-hole pairs) occur.

The analysis by Thouless \cite{Thouless} of the consequences of Eq. (\ref{Thouless}) for a long
thin wire has led to extremely important results. First, it showed that 1D
localization should manifest itself not only in ``mathematically 1D" systems but also
in the conduction in realistic, finite cross-section, thin wires, demonstrating also
the usefulness of the block-scaling point of view. Second, the understanding of the
effects of finite temperatures (as well as other experimental parameters) on the
relevant scale of the conduction, clarifies the relationships between g(L) and
experiment in any dimension. Third, defining and understanding the conductance $g(L)$
introduces the basis for the scaling theory of the Anderson localization transition \cite{Gang4}. Here, we use the results for the
(macroscopic) $T = 0$ conductivity around the localization transition to get the finite temperature conductivity there.

\subsection{The critical behavior of the $T = 0$ conductivity}
\label{critsigma}

Near, say, the lower mobility edge, $E_{m}$, the conductivity
$\sigma_0(E)$ vanishes for $E< E_{m}$ and approaches zero for $E
\rightarrow E_{m}$ from above, in the manner:

\be \sigma_0(E)= A (E - E_{m})^{x}, \label{crit} \ee
A being a
constant and $x$  the conductivity critical exponent for
localization, which has so far eluded a precise determination either
theoretically or experimentally. Within the scaling theory
\cite{Gang4}, $x$ is equal to the  critical exponent of the
characteristic length ($\xi$), because \be \sigma \sim
\frac{e^2}{\pi\hbar \xi}. \label{xi} \ee In that case, an appealing
intuitive argument by Mott \cite{Mott} and Harris \cite{Harris}
places a lower bound on $x$: \be x \geq 2/d. \label{Harris} \ee In
fact, Eq. (\ref{xi}) may be expected to hold on dimensional grounds
for any theory which does not generate another critical quantity
with the dimension of length. This should be the case for models
which effectively do not have electron-electron interactions. With
electron-electron interactions, for example, we believe that the
critical exponent for the characteristic length should satisfy an
inequality such as Eq. (\ref{Harris}) \cite{Av}. However, this may
no longer be true for the conductivity exponent.

\subsection{The conductivity at finite temperatures}

In Eqs. (\ref{current}) and (\ref{G_current}) we calculated, at
$T=0$, the current in an infinitesimal (linear response) energy
strip of width $eV$ around the Fermi energy. Generalizing this to an
arbitrary energy at finite temperature, we find that the current due
to a strip $dE$ at energy $E$ is \be I(E) dE =e
N_\ell(E_{_F})\tau^{-1}_{_L}(E) [f_{l} (E) - f_{r}(E)] dE,
\label{current(E)} \ee $f_{l}(E)$ ($f_{r}(E)$) being the Fermi
function at energy E at the left (right). The total current is
obtained by integrating Eq. (\ref{current(E)}) over energy. For
linear response $f_{l} (E) - f_{r}(E) =  eV [-\frac {\partial
f}{\partial E}]$. This gives \be \sigma(T) = \int_{-\infty}^{\infty}
dE \sigma_0 (E) [-\frac {\partial f}{\partial E}], \label{sigma(T)}
\ee where $\sigma_0 (E)\equiv
(e^2/\pi\hbar)\frac{V_{_L}(E)}{d_{_L}(E)} L^{(2 - d)}$ is the
conductivity (using Eqs. \ref{Thouless} and \ref{G_scaling}) at energy E, which would be the $T = 0$ conductivity of
the sample with $E_{F} = E$.

\subsection{Analysis of $\sigma(T,\mu-E_{m})$}
\begin{figure}[b!]
\includegraphics[width=0.5\textwidth]{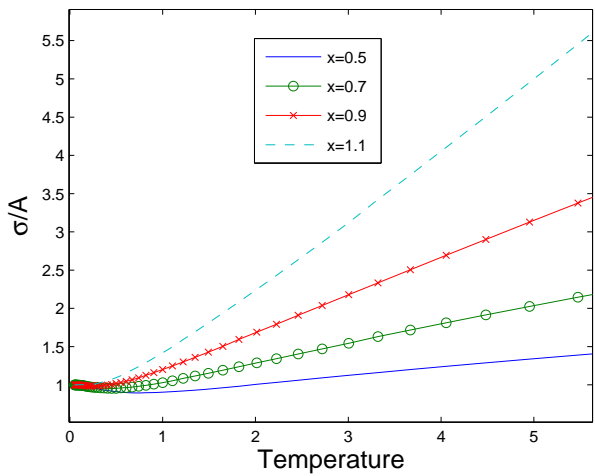}
\caption {The integral of Eq. (\ref{sigma}) was evaluated
numerically, for different values of the exponent $x$, for
$\mu-E_m=1$. Shown is the dependence of the conductivity, measured
in units of $A$ (see Eq. (\ref{crit})) as a function of temperature
(the energy scale is set by $\mu-E_m$). At low temperatures, the
conductivity saturates to a value given by Eq. (\ref{sigma_lowT}),
while at high temperatures it scales as $T^x$, see Eq.
(\ref{sigma_high_T}). The derivative at low temperatures can be
positive or negative, depending on $x$ (see Eq.
(\ref{sigma_lowT})).} \label {conductance}
\end{figure}

From now on we assume Eq. (\ref{crit}) to hold.  Measuring all
energies from the chemical potential $\mu$ and scaling them with
$T$, we rewrite Eq. (\ref{sigma(T)}) in the manner (we shall employ units
in which the Boltzmann constant, $k_{B}$ is unity, and insert it
in the final results)
\be
\sigma(T,\mu-E_{m}) = A T^{x} \Sigma ([\mu-E_{m}]/T),
\label{sigma(T,E)} \ee where the function $\Sigma(z)$ is given by:
\be \Sigma(z) \equiv \int_{-z}^{\infty} dy (y + z)^x   [-\frac
{\partial [1+ exp(y)]^{-1}}{\partial y}]. \label{sigma}\ee
Fig. \ref{conductance} shows a numerical evaluation of this
integral.

Let us consider the low and high temperature limits of this
expression, that will also be useful later for the analysis of the
thermopower.

At low temperatures, we can use the Sommefeld expansion, to obtain:
\be \sigma_{low}(T)/A=(\mu-E_m)^x+\frac{\pi^2}{6}T^2
x(x-1)(\mu-E_m)^{x-2} \label {sigma_lowT}. \ee

Notice that $\frac{\partial \sigma}{\partial T}$ is negative for
$x<1$: this comes about since in this case the function $\sigma(E)$
is concave.

At high temperatures, one can set $E_m=0$, since the contribution to
the integrals comes from energies smaller or of the order of the
temperature, and $T \gg \mu-E_m$. Therefore we have to evaluate:

\be \sigma_{high}(T)/A= \int_{0}^{\infty} E^{x} (-\frac{\partial
f}{\partial E})dE. \label {sigma_high_T} \ee

Thus, at high temperatures $\sigma_{high}/A \sim T^{x}$, with the
coefficient given by $\int_{0}^{\infty}
\frac{q^{x}}{2+\rm{cosh}(q)}dq$. In section
\ref{thermopower_section} we show that this integral can be
connected with the Riemann Zeta function, and its value is given by
Eq. (\ref{G}).

This  scaling could be used for a determination of the exponent $x$.
However a much closer determination of that exponent would follow
from the scaling of both the conductivity and the thermal and
thermoelectric transport coefficients, which will be studied in the
coming sections.

\section{Thermal and thermoelectric transport}
\label{therm}

\subsection{General relationships}
Consider now the case where both a voltage $V$ and a temperature
difference $\Delta T$ are applied between the two blocks. We choose
for convenience $k_B \equiv 1$. Both are small enough for linear
response to hold. Here we have to replace the Fermi function
difference in Eq. (\ref{current(E)}) by
$$f_{l} E) - f_{r}(E) =  eV [-\frac {\partial f}{\partial E}]
+ \Delta T [-\frac {\partial f}{\partial T}].$$ Then generalizing
Eq. (\ref{sigma(T)}) yields for the electrical current \be I = \int
dE \frac{G(E)}{e} \{[-e\frac {\partial f}{\partial E}] V +  [-\frac
{\partial f}{\partial T}] \Delta T\}, \label{Ie} \ee with $G$ given
in Eq. (\ref{G_scaling}). The first term is the ordinary ohmic
current and the second one is the thermoelectric charge current due
to the temperature gradient.

Next, we derive the heat current. The heat carried by an electron
with energy $E$ (measured from the chemical potential, $\mu$) is
equal to  $E$. This is shown, for example in
Ref. [\refcite{Ziman}] by noting that the heat is the difference
between the energy and the free energy. Sivan and Imry \cite{SI}
verified it in their Landauer-type model by calculating the flux of
$TS$ along the wire connecting the two reservoirs. In the appendix,
we obtain the same result in our block model, from the time
derivative of the entropy of each block. Thus, we obtain the heat
current $I_{Q}$,
\be I_{Q} = \int E dE  \frac{G(E)}{e^2}
\{[-e\frac{\partial f}{\partial E}] V +  [-\frac{\partial f}{\partial
T}] \Delta T\}. \label{IQ} \ee
Here the first term is the
thermoelectric heat current due to the voltage, while the second one
is the main contribution to the usual electronic thermal
conductivity $\kappa$.

In this model the ratio of {\it   thermal} to electrical conductivities is of the order of $(k_{B}/e)^{2} T$. This is because
a typical transport electron carries a charge $e$ and an excitation energy
of the order of $k_{B} T$ and the driving forces are the differences in $eV$ and $k_{B} T$. This ratio is basically the Wiedemann-Franz law.

It is convenient to summarize Eqs. (\ref{Ie}) and (\ref{IQ}) in
matrix notation \cite{Callen}: \be \left(
\begin{array}{ccc}
  I  \\
  I_{Q}\\
\end{array}
 \right)=
\left(
\begin{array}{ccc}
 L_{11} & L_{12}\\
 L_{21}&L_{22}\\
 \end{array}%
\right)
\left(
\begin{array}{ccc}
  V \\
  \Delta T\\
\end{array}
 \right), \label{matrix}
\ee where the coefficients $L_{ij}$ can be read off Eqs. (\ref{Ie})
and (\ref{IQ}). Since $f$ is  a function of $E/T$, we see that \be
-\frac {\partial f}{\partial T} =\frac{E}{T} \frac {\partial
f}{\partial E} \label{deriv} \ee Therefore, the two ``nondiagonal"
thermoelectric coefficients: the one relating $I$ to $\Delta T$,
$L_{12}$, and the one relating $I_{Q}$ to $V$, $L_{21}$, are equal
within a factor T. \be L_{12} = L_{21}/T. \label{Onsager0} \ee This
is an Onsager \cite{Callen,Onsager,Ziman,LL} relationship, which holds very
generally for systems obeying time-reversal symmetry (and particle
conservation -- unitarity). The case where time-reversal symmetry is
broken, say by a magnetic field, is briefly discussed in the next
subsection.

We conclude this subsection by defining and obtaining an expression
for the absolute thermoelectric power (henceforth abbreviated as
just ``thermopower") of a material. Suppose we apply a temperature
difference $\Delta T$ across a sample which is open circuited and
therefore no current can flow parallel to $\Delta T$. To achieve
that, the sample will develop a (usually small) voltage $V$, so that
the combined effect of both $\Delta T$ and V will be a vanishing
current. From Eqs. (\ref{Ie}), (\ref{matrix}) and (\ref{deriv}) we
find that the ratio between $V$ and $\Delta T$, which is defined as
the thermopower, S, is given by \be S \equiv \frac{V}{\Delta T}
=-\frac{L_{12}}{L_{11}} =  \frac{\int dE E \sigma_0(E) \frac
{\partial f}{\partial E}} {eT \int dE \sigma_0(E) \frac {\partial
f}{\partial E}} \label{S}. \ee

\subsection{Onsager relations in a magnetic field}

From time-reversal symmetry at $H = 0$ and unitarity (particle conservation) follows the Onsager relation
\cite{Callen,Onsager,Ziman,LL} for the $T = 0$ conductance {\it }
\be
\sigma(E,H)= \sigma(E,-H). \label{Onsagersigma}
\ee
This can be proven for our model from the basic symmetries of the interblock matrix elements. This symmetry obviously follows for the temperature-dependent electrical and thermal conductivities $\sigma(T)$ and $\kappa(T)$.

For the nondiagonal coefficients, the usual Onsager symmetry reads
\be L_{12}(H) = L_{21}(-H)/T. \label{Onsager-H} \ee In our case,
since the nondiagonal coefficients are expressed as integrals over a
symmetric function (Eq. (\ref{Onsagersigma})), they also obey \be
L_{ij}(H) = L_{ij}(-H). \label{OnsagerH} \ee {\it i.e.} the
nondiagonal coefficients are symmetric in $H$ as well.

\subsection{Analysis of the thermopower}
\label{thermopower_section}

Eq. (\ref{S}) for the thermopower  is identical to  the one
derived in two-terminal linear transport within the Landauer
formulation in Ref. [\refcite{SI}], which is equal in the
appropriate limit to the Cutler-Mott \cite{CM,Mott} expression:
\begin{equation}
S =  \frac{ \int_{E_m}^{\infty} dE (E - \mu) \sigma_0(E)
(-\frac{\partial f}{\partial E})}{e \sigma(T) T}, \label{thermo}
\end{equation}
where $\mu$ is the chemical potential, $\sigma_0(E)$ is the
conductivity for carriers having energy $E$ and $\sigma$ is the
total conductivity. The Physics of this formula is clear for the
(Onsager-dual) Peltier coefficient: a carrier at energy $E$ carries
an excitation energy (similar to heat, see the appendix) of $E - \mu$.

Clearly, electrons and holes contribute  to $S$ with opposite signs.
S will tend to vanish with electron-hole symmetry and will be small,
as happens in many metals, especially in ordered ones, when the
variation in energy of $\sigma_0(E)$ around $\mu$ is weak.

Having a strong  energy dependence of $\sigma_0(E)$, and being very
different above and below $\mu$ will
 cause relatively large values of $S$. We believe that this is what happens in disordered narrow-gap semiconductors,
 which feature in many present-day good thermoelectrics. As noted in Refs. [\refcite{CM,SI}], the
 Anderson metal-insulator transition (or at least its vicinity) offers an almost ideal situation for large thermopowers.
There, $\sigma_0(E)$ vanishes below the mobilty edge $E_{M}$ (for
electrons) and approaches zero,
 probably with an infinite slope, above it. Hopping processes in the localized phase are not considered here. A brief analysis in Ref. [\refcite{SI}] demonstrated that S
 scales with $ z \equiv(\mu - E_{M} )/T$:
 \be
 S = Y(\frac{\mu - E_{M}}{T}), \label{scaling}
 \ee
 (Y being a universal scaling function)
 and assumes the two limits:
\begin{equation}
S \sim (\mu - E_{m})^{-1},~~ for~~z \gg 1); \label{result}
\end{equation}
and
\be
S \sim const - z,~~ for~~z \ll 1. \label{result1}
\ee

Of course, there is no ``real" divergence of $S$ \cite{Enderby},
since when ($\mu - E_{m}) \rightarrow 0$ (and the slope of $S(T)$
diverges), it will eventually become smaller than $T$ and the
large-slope linear behavior  will saturate as in Eq.
(\ref{result1}).

\begin{figure}[b!]
\includegraphics[width=0.5 \textwidth]{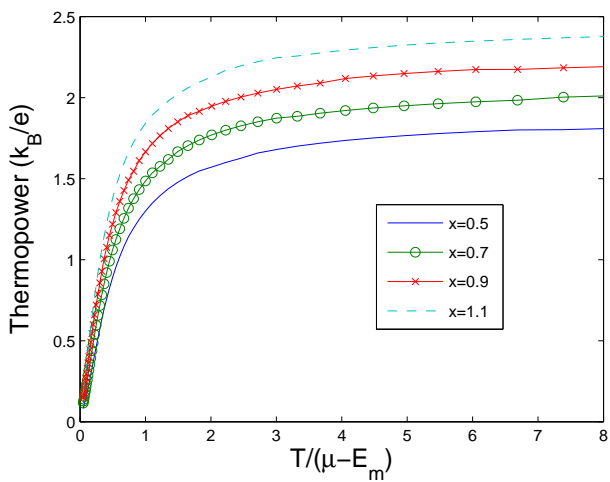}
\caption {The integrals of Eq.\ (\ref{thermo}) were evaluated
numerically, for different values of the exponent $x$. At low
temperatures, the thermopower is linear with temperature, while at
high temperatures it saturates, to a value which depends on $x$.} \label
{thermopower_numeric}
\end{figure}

Fig. \ref{thermopower_numeric} shows a numerical evaluation of Eq.
(\ref{thermo}), demonstrating the linear low temperature regime and
the saturation at high temperatures. Let us now make a more thorough
investigation of the low and high temperature regimes.

For low temperatures, we can use, as before, the Sommerfeld
expansion for the nominator and denominator, to obtain:

\be S_{low} \approx \frac{\frac{\pi^2}{3}T x (\mu-E_m)^{x-1}
+O(T^3)}{e[(\mu-E_m)^x+\frac{\pi^2}{6}T^2
x(x-1)(\mu-E_m)^{x-2}+O(T^4)]}\label {slope_theory} \ee

A more complete expression is give in Eq. (\ref{full_som}) below.

Thus, at $T \ll \mu- E_m$, the thermopower is linear in temperature:

\be S_{low} \approx \frac{\pi^2x T }{3e (\mu-E_m)} +O(T^3). \label{low} \ee

\begin{figure}[b!]
\includegraphics[width=0.5\textwidth]{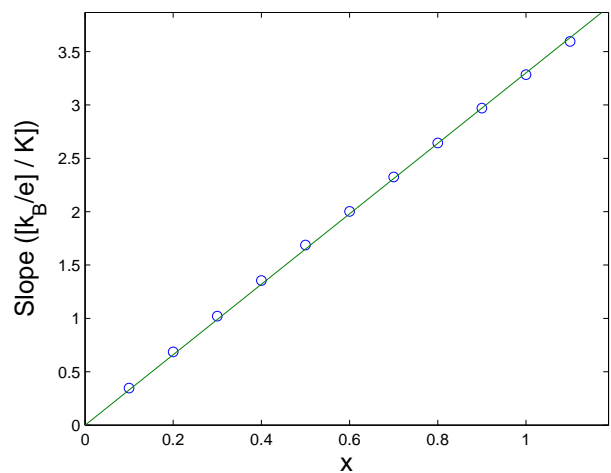}
\caption {At low temperatures, the thermopower is described by Eq.\
(\ref{slope_theory}). The plot shows a comparison between this
expression and the evaluation of the slope extracted from the
numerics demonstrated in  Fig. \ref{thermopower_numeric}.} \label
{slope}
\end{figure}

\noindent Fig. \ref{slope} compares this expression with the numerically
evaluated slope.

At high temperatures, one can set $E_m=0$, as before. Therefore we
have to evaluate:

\be S_{high}= \frac{\int_{0}^{\infty} E^{x+1} (-\frac{\partial
f}{\partial E})dE}{eT \int_{0}^{\infty} E^{x} (-\frac{\partial
f}{\partial E})dE}.\ee

\noindent We can write $\int_{0}^{\infty} E^{\beta} (-\frac{\partial
f}{\partial E})dE= T^{\beta+2}G(\beta)$, with the dimensionless
function $G(\beta)$ defined as:

\be G(\beta)= \int_{0}^{\infty}  \frac{m^{\beta}}{2+\rm{cosh}(m)}dm.
\label{G_int} \ee

\noindent We then have $S_{high}=\frac{G(x+2)}{eG(x+1)}$.

In fact, the integral of Eq. (\ref{G_int}) can be related to the
Riemann Zeta function $\zeta$:

\be \zeta(\beta)=\frac{1}{\Gamma(\beta)} \int_{0}^{\infty} \frac{
m^{\beta-1}}{e^m-1}dm. \ee

\noindent Defining $C= \int_{0}^{\infty} \frac{ m^{\beta-1}}{e^m-1}dm$, we
find that:

\be C- G/\beta = \int_{0}^{\infty} \frac{
2m^{\beta-1}}{e^{2m}-1}dm=C/2^\beta, \ee

\noindent therefore:

\be G(\beta)= \beta C (1-1/2^{\beta-1})=\beta \zeta(\beta)
\Gamma(\beta) (1-1/2^{\beta-1}). \label{G} \ee

\noindent This gives an exact formula for the thermopower at high temperatures:

\be S_{high}=(1+x)\frac{\zeta(1+x)(2^{x}-1)}{e\zeta(x)(2^{x}-2)}.
\label {high_exact}\ee

\noindent At $x=0$, one obtains $S=2\rm{log}(2)$, while for $x \gg 1$, one
obtains $S_{high} \approx 1+x$.

Actually, understanding the behavior for large $x$ is simple: If we
were to approximate the derivative of the Fermi-function by
$e^{-E/T}$, we would have $G(\beta)=\Gamma(\beta)$, where $\Gamma$
stands for the Gamma function. Then, by its properties, we immediately have
that $S_{high} \approx(1+x)/e .$

It turns out that a good approximation to $S_{high}(x)$ can be
obtained by interpolating the exact $x=0$ result and the large $x$
result, by the form:

\be S_{high} \approx \frac{1}{e} [2 \rm{log}(2) +x]. \label{high_T}
\ee

\begin{figure}[b!]
\includegraphics[width=0.5\textwidth]{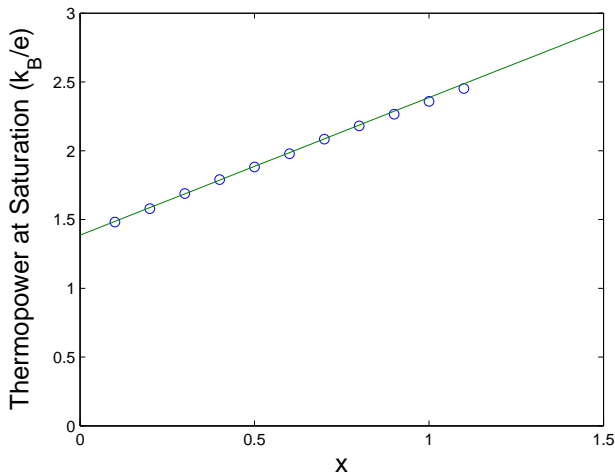}
\caption {At high temperatures, the thermopower saturates, at a
value dependent on $x$. The plot shows the saturation values
extracted by evaluating Eq.\ (\ref{thermo}) numerically (see Fig.
\ref{thermopower_numeric}), and a linear dependence corresponding to
Eq.\ (\ref{high_T}). \label {saturation}}
\end{figure}

\noindent Fig. \ref{saturation} compares the exact saturation values of Eq.
(\ref{high_exact}) with this approximate form. We found that the
difference for all values of $x$ is less than 6 percent, and
therefore Eq. (\ref{high_T}) provides a practical working formula
for the saturation value of the thermopower.

An interesting feature of the crossover from the low to high
temperature regime, is the possibility of an inflection point in the
thermopower dependence. Similar to the behavior of the conductance,
which grew for $x>1$ but diminished for $x<1$, here there will be an
inflection point for $x<1$. To see this, we have to calculate the
next order in the Sommerfeld expansion in the nominator $Q$ of Eq.
(\ref{slope_theory}), to obtain, up to corrections of order
$O(T^5)$:

\be Q=\frac{\pi^2}{3}T x (\mu-E_m)^{x-1} +\frac{7\pi^4}{90}T^3
x(x-1)(x-2) (\mu-E_m)^{x-3}. \ee

This leads to the following low temperature correction of the
thermopower:

\be S_{low} \approx \frac{\pi^2x T }{3e (\mu-E_m)}+
\frac{\pi^4}{45}x(x-1)(x-7)\frac{T^3}{e(\mu-E_m)^3} +O(T^5) \label{full_som},\ee

implying an inflection point for $x<1$.

\section{Brief discussion of experiments}
\label{exp}

Large thermopowers that are linear in the temperature, at least in the
metallic regime, were already
found in the pioneering extensive work on Cerium sulfide compounds by Cutler
and Leavy \cite{CL}, and analyzed by Cutler and Mott \cite{CM}. It is
interesting to address specifically the behavior around the localization
transition. An experiment performed on $In_2O_{3-x}$ (both amorphous and
crystalline), approaching the Anderson MIT, shortly after  Ref.
[\refcite{SI}], confirmed qualitatively the main features of Eqs.
(\ref{result}) and (\ref{result1})\cite{Zvi}.  Values of S exceeding
$100 \frac{\mu V}{K}$ were achieved. It should be kept in mind that
for a good determination of the critical exponent $x$, one needs
data at low temperatures and small $\mu-E_m$. Data too far from the
QPT, which is at both $T = 0$ and $\mu-E_m = 0$, will not be in the
critical region and may be sensitive to other effects, as will be discussed
later.

It has been customary to use only the low temperature conductivity
to determine the critical exponent $x$. Using similar $In_2O_{3-x}$
samples,  the conductivity was extrapolated in Ref.
[\refcite{Zvi1}] to $T = 0$ and those values were plotted against
a control parameter which should be proportional to $\mu-E_m$ when
both are small. A value of $x = .75-.8$ was found.

\begin{figure}[b!]
\includegraphics[width=0.5\textwidth]{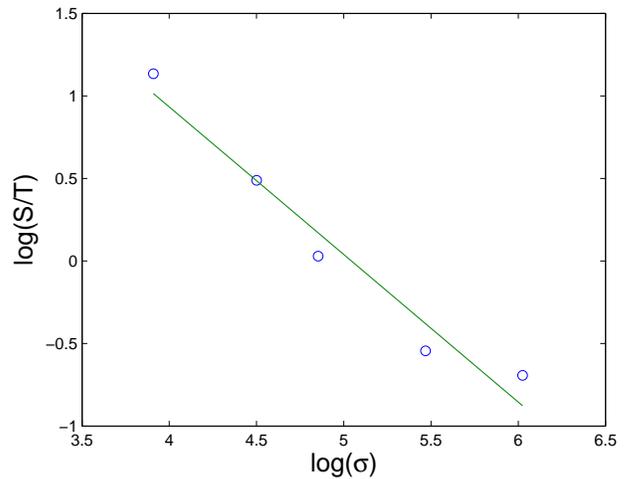}
\caption { A small set of data, which should be taken as a
preliminary to more extensive studies, was used to find the critical
exponent $x$. Eqs. (\ref{crit}) and (\ref{result}) show that the
linear (low-temperature) regime of the thermopower has a slope which
has an inverse power-law dependence on the distance from the
transition. The slope of the
log-log plot gives $x \sim 1.1$, with a significant error.
 } \label {s_sigma}
\end{figure}

It would be much better to use both the above conductivity
values and the slopes of $S(T)$ near $T = 0$, according to Eq. (\ref{low}). An even
 better way to do that would be to eliminate the control parameter $\mu-E_n$ from
 Eqs. (\ref{crit}) and (\ref{result}), getting
 \be
 \frac{dS}{dT}  _{T\rightarrow 0} \sim [ \sigma(T = 0)]^{-1/x} ,\label{direct}
 \ee
not having to determine the additional parameter $\mu-E_m$ for each
case. The data allowed us to effect this only approximately, see
Fig. \ref{s_sigma}, giving $x \cong 1 \pm .2$. However getting near
the QPT, this procedure appears to be the one of choice.

 Obviously, using the two full functions $\sigma (\mu-E_m,T)$ and  $S (\mu-E_m,T)$,
 in the critical region  would place even more strict constraints on $x$. Below, we do this for
 the existing data, to demonstrate the method. Their scaling works well,
 but the value of $x$ obtained is not likely to be the real critical value.
 This is due to a few caveats which will be mentioned.

\begin{figure}[b!]
\includegraphics[width=0.5\textwidth]{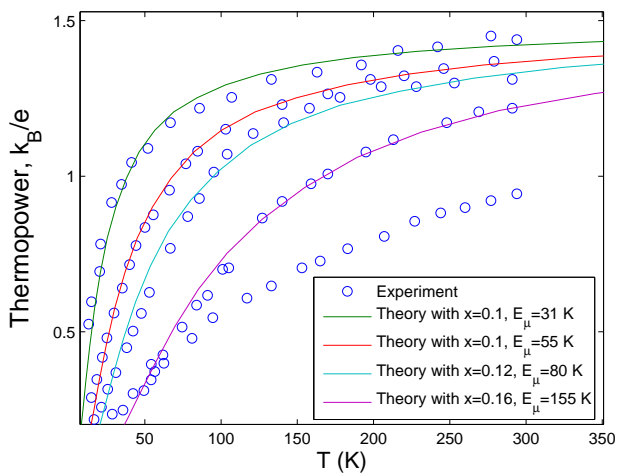}
\caption {A fit of the theoretical expression of Eq. (\ref{thermo})
to the experimental data. The data was taken from  Ref.
[\refcite{Zvi}]. The thermopower is measured in units of
$k_B/e\sim 86 \mu V /K $. } \label {zvi_fit}
\end{figure}

\begin{figure}[b!]
\includegraphics[width=0.5\textwidth]{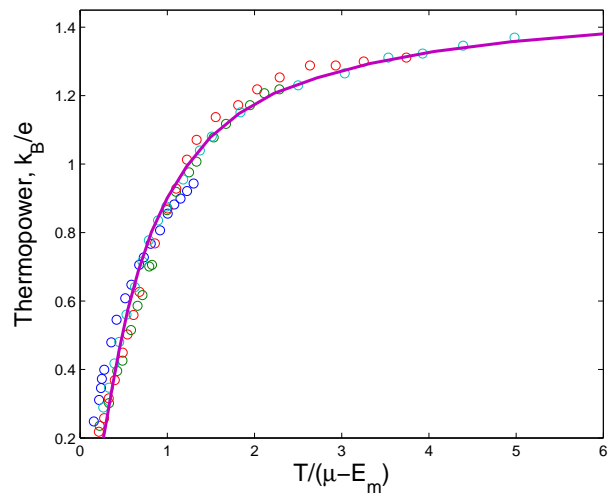}
\caption {The 4 data sets of Fig. \ref{zvi_fit} which are closest to
the transition are shown, scaled to lie on a single universal curve,
as function of $T/E_{m}$. x was taken as $0.1$ and each curve was
given a single value of $E_{m}$. } \label {thermopower_collapse}
\end{figure}

Fig. \ref{zvi_fit} compares the above predictions to the
experimental data, taking the exponent $x$ and the values of $E_{m}$
as fitting parameters.  Fig. \ref{thermopower_collapse} shows the
approximate data collapse obtained by rescaling the temperature axis
of each of the measurements (corresponding to the appropriate value
of $E_M$) as in Eq. (\ref{scaling}), and the theoretical curve
corresponding to $x=0.1$.

~

The fit is certainly acceptable. However, the value of $x =.1$ is
both in disagreement with the previously determined value and
impossible for noninteracting electrons, since there $x > 2/3$ in
three dimensions. Although, as explained at the end of subsection
\ref{critsigma}, this constraint may not be valid with interactions,
we do not take this last value of $x$ seriously. Since the
saturation value was shown be approximately given by a $x+C$, with
$C \sim 1.39$ (see Eq. (\ref{high_T})), a change of the thermopower
by tens of percents will cause a large change in the deduced value
of $x$. These last fits should be regarded only as demonstrating our
recommendations for a possible extension of the analysis of future
experimental studies.

At higher temperatures, the analysis will be influenced  not only by
data that are not in the critical region, but two further relevant
physical processes may
well come in. Obviously, inelastic scattering (by both phonons and other
electrons) will be more important. Moreover, for $T \gtrsim \mu - E_{m}$,
some of the transport will occur via hopping  of holes below the mobility
edge. Their
thermopower might cancel some of the contribution of electrons above $\mu$
and thus reduce the thermopower below the values considered here. This
clearly needs further treatment.

At any rate, the interactions appear to be strongly relevant and may
give unexpected values for $x$. Measurements at lower temperatures
and closer to the transition are clearly needed. Checking
simultaneously the behavior of both the conductivity and the
thermopower is suggested as the method of choice for this problem.

It must be mentioned that
similar experiments on granular Al did not show the expected
behavior. It should be kept in mind that the resistivities needed to
approach the MIT for $\sim 100A$ grains are larger \cite{book} than
the ones for microscopic disorder. Another relevant issue, which we
are going to examine in detail in future work, is that while $E_{F}$  is in the $eV$ range, all the energies (without
electron-electron interactions) relevant for localization are
smaller by several orders of magnitude than for microscopic
disorder. Thus, the temperature range for the enhanced thermopower
might well be in the sub $K$ range, which was not addressed in the
experiments. The Coulomb blockade may partially alleviate this, but
only when it is operative (not in the metallic regime).

The later
experiment of Ref. [\refcite{Exp}], on $Si:P$ obtained very
modest enhancement of S, but were stated to have been dominated by
effective magnetic impurities, which are known to be strongly
relevant for the Anderson transition (e.g. eliminating the weak
localization contributions). All these issues have to be clarified.

\section{Concluding remarks}
\label{Disc}

In this paper we have shown how to include the thermal current in
the Thouless scaling picture of conduction in  disordered systems.
Expressions were given for the $2\times2$ matrix of longitudinal
thermoelectric coefficients, in terms of $\sigma_0(E)$, the $T=0$
conductivity of the system were its Fermi level fixed at the energy
$E$. The Onsager relations were shown to hold within this
formulation. For the usual critical behavior of $\sigma_0(E)$, given
by Eq. (\ref{crit1}), these behaviors were analyzed for an arbitrary
ratio of $T$ to the distance to the mobility edge. They were shown
to satisfy scaling relationships which were confirmed numerically
along with their limiting behaviors.

It was shown how the conductivity and thermopower data close to the
Anderson QPT should be analyzed simultaneously to yield a better
estimate of the critical exponent $x$ than the determination based
on $\sigma(T)$ alone. This was done for the low-temperature limits
of existing data \cite{Zvi,Zvi1} on the transition in $In_2O_{3-x}$,
giving already a good ballpark estimate of $x$.  The data going to
higher (probably too high)  temperatures do scale and collapse
according to Eq. (\ref{scaling}), but the resulting values of $x$
appear to be too small.  One may speculate that this is due to
interaction effects, but we prefer to postpone this
to after having done this analysis with lower temperature data closer to the transition.

 Similar experiments on granular Al do not produce a large and  interesting thermopower as above. This is certainly a
 matter for concern. The explanation might well be due to the smaller microscopic conductivity scales ($\frac {e^{2}}{\hbar R},
 R$ being the grain size), or to the different energy scales
 relevant for these systems \cite{book}. Alternatively, the inelastic scattering, not treated in this paper, may be relevant as well.

 The sharp and asymmetric behavior of $\sigma_0(E)$ near the transition
is ideal for getting large thermopowers. The predicted values approach $\sim 200 \frac{\mu V}{K}$. While the experimental
results \cite{Zvi} on $In_2O_{3-x}$ are smaller by $\sim 40 \%$, this is still encouraging.
Were it possible to increase these values say by phonon drag effects,  this might even become  applicable.
Clearly, a treatment of the effects of inelastic scattering on the thermopower is called for,
especially including  the hopping conductivity regime.

\appendix 
\section{The heat carried by a transport quasiparticle}

To make this analysis useful also for heat transport by phonons,
etc., we display the equations for both fermions and bosons. The
entropy associated with a state of a given equilibrium system at energy $E$,
having a population $f$ is \be S_{E} = -k_{B} [f ln f + (1\pm f) ln
(1 \pm f)], \label{entr} \ee where the upper (lower) sign is for
bosons (fermions). When the population f changes with time, the
change of $S_{E}$ with time is \be \dot S_{E} = -k_{B} \dot f ln
\frac{f}{1 \pm f} = -\frac{E}{T} \dot f, \ee where to get the last
equality we used the equilibrium $f = (e^{\frac{E}{T}} \mp 1)^{-1}$.
The outgoing heat current  $T \dot S$ is the time derivative of the
population times the excitation energy. Thus, each particle leaving
the system carries ``on its back" an amount of heat $E$ which is its
energy (measured from $\mu$).
Summing $\dot S_{E}$ over all energies shows that the outgoing heat
current is given by the outgoing particle current where the
contribution of each energy is multiplied by $E - \mu$.

It should be noted
that the equality of the amounts of $E-\mu$ and $-TS$ carried by the excitation implies
that the  relevant free energy  does not change when the (quasi)particle moves to another
system which is  in equilibrium with the first one. This is true for equilibrium fluctuations
and also for linear response
transport ($V \rightarrow 0$ and $\Delta T \rightarrow 0$), between the two systems.

As remarked, the result that the heat carried by an electron is given by its  energy
measured from the chemical potential, $\mu$,
is valid also for bosons. As a small application, one can easily calculate the net thermal current
carried by a single-mode phonon/photon waveguide fed by thermal baths at $T \pm \Delta T /2$.
The result is a thermal conductance of $k_B^{2} T \pi/(6 \hbar)$ (per mode),
with no reflections. This agrees with the result of [\refcite{Kir}]. The sound/light velocity cancels
between the excitation velocity and its (1D) DOS, exactly as in the electronic case. This is why
this result and the one based on the Wiedemann-Franz law for electrons are of the same order of magnitude.
That their numerical factors are equal is just by chance. With reflections due to disorder,
once the waveguide's
length is comparable to or larger than the localization length (mean free path for a single mode),
its thermal conductance drops markedly.

\vspace{.5cm}

\section*{Acknowledgments}
We thank Uri Sivan, Ora Entin-Wohlman,  Amnon Aharony and the late C. Herring for discussions. Special thanks are due to Zvi Ovadyahu for instructive discussions and for making his data available to us.
This work was supported by the German Federal Ministry of
Education and Research (BMBF) within the framework of the
German-Israeli project cooperation (DIP), by the Humboldt
Foundation, by the US-Israel
Binational Science Foundation (BSF), by the Israel Science
Foundation (ISF) and by its Converging Technologies Program.
YI is grateful to the Pacific Institute of Theoertical Physics (PITP) for its hospitality when some of this work was done.


\begin{thebibliography}{999}
\bibitem{small} Were one lucky to have very weak interaction in some system, the crossover region to the
interaction-dominated behavior would be small and approximate information on the noninteracting critical behavior could be obtained from experiments on this system.
\bibitem{lrsymm} This symmetry is valid also when there is no symmetry between the blocks (say, when $N_{l}(E) \neq N_{r}(E)$).
We shall however mainly consider here the case where the blocks are of the same size ($N_{l}(E) = N_{r}(E)$), but they usually do have different defect configurations.
\bibitem{Av}
This is because this inequality assures that the characterisic length, $\xi$, grows quickly enough near the transition to average out the disorder fluctuations between different volumes of $\xi^{d}$. This implies that no intrinsic inhomogeneity is generated when the transition is approached.
\end{thebibliography}
\end{document}